TABLE 1
Numerical Values[a]

| $(l_e, l_m)$ | | | | | | | | | (10.8, 2) | (20.1, 21) | (52.8, 56) | (55.8, 55) |
|---|---|---|---|---|---|---|---|---|---|---|---|---|
| $(l_{e^{-0.5}})$ | | | | | | | | | ((2), 25) | (14, 29) | (38, 77) | (37, 76) |
| $\sqrt{I(W_l)}$ | | | | | | | | | 1.62 | 1.02 | 1.02 | 1.15 |

| # | $(\Omega_0, h, \Omega_B h^2)$ | $t_0$ (Gyr) | $S$[b] | $(l_{p1}, l_{p2}, l_{p3})$[d] | $Q_{\rm rms-PS}^{\rm (no-Q)}$[e] ($\mu$K) | $AVh^4$[f] ($10^6$Mpc$^4$) | $\frac{\delta M}{M}\|_{8h^{-1}{\rm Mpc}}$ | $\beta_I$[h] | FIRS[j] ($\mu$K) | TENERI[l] ($\mu$K) | BARTOL[m] ($\mu$K) | SK94K3[n] ($\mu$K) |
|---|---|---|---|---|---|---|---|---|---|---|---|---|
| (1) | (2) | (3) | (4) | (5) | (6) | (7) | (8) | (9) | (10) | (11) | (12) | (13) |
| 1 | (0.1, 0.75, 0.0125) | 11.7 | **0.059** | (778, 1910, 2960) | 23.1 | 0.795 | **0.086 – 0.11** | **0.028 – 0.035** | 29 – 37 | 25 – 31 | 22 – 28 | **22 – 28** |
| 2 | (0.2, 0.65, 0.0175) | 12.7 | **0.10** | (544, 1330, 2040) | 26.5 | 0.946 | **0.20 – 0.25** | **0.10 – 0.12** | 31 – 38 | 26 – 32 | 25 – 32 | **26 – 32** |
| 3 | (0.2, 0.70, 0.0125) | 11.8 | **0.12** | (522, 1250, 1920) | 26.5 | 0.944 | **0.25 – 0.32** | **0.13 – 0.16** | 31 – 38 | 26 – 32 | 25 – 31 | **25 – 32** |
| 4 | (0.2, 0.75, 0.0075) | 11.0 | **0.14** | (509, 1180, 1830) | 26.5 | 0.945 | **0.31 – 0.39** | **0.15 – 0.19** | 31 – 38 | 26 – 32 | 25 – 31 | **25 – 31** |
| 5 | (0.3, 0.60, 0.0175) | 13.2 | **0.15** | (434, 1050, 1610) | 25.8 | 1.04 | 0.35 – 0.43 | 0.22 – 0.27 | 30 – 38 | 27 – 34 | 29 – 36 | **29 – 36** |
| 6 | (0.3, 0.65, 0.0125) | 12.2 | **0.17** | (417, 989, 1520) | 25.8 | 1.04 | 0.43 – 0.53 | 0.27 – 0.34 | 30 – 38 | 27 – 34 | 28 – 35 | **29 – 36** |
| 7 | (0.3, 0.70, 0.0075) | 11.3 | 0.20 | (407, 937, 1440) | 25.8 | 1.04 | 0.51 – 0.64 | 0.32 – 0.40 | 30 – 38 | 27 – 34 | 28 – 34 | **28 – 35** |
| 8 | (0.4, 0.60, 0.0175) | 12.7 | 0.20 | (362, 869, 1320) | 23.4 | 1.10 | 0.54 – 0.68 | 0.41 – 0.51 | 30 – 37 | 28 – 35 | 32 – 39 | 32 – 40 |
| 9 | (0.4, 0.65, 0.0125) | 11.7 | 0.23 | (349, 822, 1250) | 23.4 | 1.09 | 0.65 – 0.81 | 0.49 – 0.61 | 30 – 37 | 28 – 35 | 31 – 38 | 31 – 39 |
| 10 | (0.4, 0.70, 0.0075) | 10.9 | 0.27 | (343, 779, 1200) | 23.4 | 1.09 | 0.76 – 0.94 | 0.57 – 0.71 | 29 – 37 | 28 – 35 | 30 – 37 | **30 – 37** |
| 11 | (0.5, 0.55, 0.0175) | 13.4 | 0.23 | (325, 780, 1180) | 20.5 | 1.12 | 0.65 – 0.82 | 0.56 – 0.70 | 29 – 36 | 30 – 37 | 34 – 43 | 35 – 44 |
| 12 | (0.5, 0.60, 0.0125) | 12.3 | 0.27 | (313, 736, 1120) | 20.5 | 1.12 | 0.78 – 0.98 | 0.67 – 0.84 | 29 – 36 | 29 – 37 | 33 – 41 | 34 – 42 |
| 13 | (0.5, 0.65, 0.0075) | 11.3 | **0.31** | (307, 696, 1070) | 20.5 | 1.12 | 0.91 – 1.1 | 0.78 – 0.98 | 29 – 36 | 29 – 36 | 32 – 40 | 32 – 40 |
| 14 | (1.0, 0.50, 0.0125) | 13.0 | **0.45** | (217, 505, 764) | 19.8 | 0.785 | 1.2 – 1.5 | **1.5 – 1.9** | 29 – 36 | 31 – 38 | 37 – 46 | 37 – 47 |
| Observ. | ... | ... | 0.2 – 0.3[c] | ... | ... | ... | 0.4 – 1.2[g] | 0.25 – 0.80[i] | 23 – 39[k] | 24 – 50[l] | ... | 39 – 69[n] |
| Observ. | ... | ... | ... | ... | ... | ... | ... | ... | 31[k] | 34[l] | ... | 50[n] |

[a] Bandtemperature $\delta T_l$, unless noted otherwise. The ranges on the theoretical predictions are $\pm 1\sigma$ (11% from the DMR normalization), large-scale structure observation ranges are indicative of the current status, and CMB observation ranges are either $\pm 1\sigma$ or $2\sigma$ upper limits. [b] Shape parameter $S = \Omega_0 h \exp[-\Omega_B(\Omega_0+1)/\Omega_0]$ (S95). [c] (E.g., Peacock & Dodds 1994). [d] Multipoles of the first three peaks in the $C_l$ spectrum. [e] (GRSB). [f] Energy-density perturbation power spectrum normalization factor (S95). [g] For a reasonable value of the bias parameter. [h] $\beta_I = 1.3\Omega_0^{0.6}\delta M/M(8h^{-1}{\rm Mpc})$. [i] (Cole, Fisher, & Weinberg 1995). [j] (Ganga et al. 1994). [k] For an $n = 1$ power-law ($\Omega_0 = 1$) model (not fiducial CDM), with 7% calibration uncertainty (Bond 1995). [l] For an $n = 1$ power-law ($\Omega_0 = 1$) model (not fiducial CDM), with 10% calibration uncertainty (Hancock et al. 1995). [m] (Piccirillo et al. 1995). [n] SK94K$n$ and SK94Q$n$ are the SK94 Ka and Q band $n$-point $W_l$ (Netterfield 1995). Observational numbers are from a reanalysis of the data (Netterfield et al. 1996), and account for 14% calibration uncertainty.

| (57.2, 64) | (59.4, 55) | (63.6, 54) | (66.2, 73) | (67.0, 69) | (74.4, 95) | (76.5, 75) | (80.6, 75) | (82.3, 59) | (82.9, 73) | (91.7, 73) | (96.0, 95) | (97.6, 109) | (100, 95) |
|---|---|---|---|---|---|---|---|---|---|---|---|---|---|
| (35, 98) | (37, 76) | (37, 76) | (40, 112) | (47, 95) | (28, 118) | (57, 96) | (57, 97) | (39, 84) | (54, 96) | (53, 99) | (76, 116) | (60, 168) | (76, 117) |
| 1.08 | 1.22 | 1.28 | 1.23 | 1.07 | 1.38 | 0.881 | 0.930 | 1.66 | 1.02 | 1.34 | 0.727 | 0.551 | 0.772 |
| SP94Ka[o] | SK94Q3[n] | SK95C3[p] | SP94Q[o] | SK93[q] | PYTHG[r] | SK94K4[n] | SK94Q4[n] | BAM[s] | SK95C4[p] | PYTHL[t] | SK94K5[n] | ARGO[u] | SK94Q5[n] |
| ($\mu$K) | ($\mu$K) | ($\mu$K) | ($\mu$K) | ($\mu$K) | ($\mu$K) | ($\mu$K) | ($\mu$K) | ($\mu$K) | ($\mu$K) | ($\mu$K) | ($\mu$K) | ($\mu$K) | ($\mu$K) |
| (14) | (15) | (16) | (17) | (18) | (19) | (20) | (21) | (22) | (23) | (24) | (25) | (26) | (27) |
| 23 – 29 | 22 – 28 | 23 – 28 | **23 – 29** | 22 – 28 | **23 – 29** | 23 – 28 | 23 – 28 | 23 – 29 | 23 – 29 | **23 – 29** | 23 – 29 | **24 – 30** | 23 – 29 |
| 26 – 33 | 26 – 33 | 27 – 33 | 27 – 34 | 26 – 33 | **28 – 35** | 27 – 34 | 27 – 34 | 28 – 35 | 28 – 35 | **29 – 36** | 29 – 36 | 30 – 37 | 29 – 36 |
| 26 – 32 | 26 – 32 | 26 – 32 | 26 – 33 | 26 – 33 | **27 – 34** | 26 – 33 | 27 – 33 | 27 – 34 | 27 – 34 | **28 – 35** | 28 – 35 | 29 – 36 | 28 – 35 |
| 25 – 32 | 25 – 31 | 25 – 32 | 26 – 32 | 25 – 31 | **26 – 33** | 26 – 32 | 26 – 32 | 27 – 33 | 26 – 33 | **27 – 33** | 27 – 33 | 27 – 34 | 27 – 34 |
| 30 – 37 | 30 – 37 | 30 – 38 | 31 – 38 | 30 – 38 | **32 – 40** | 31 – 39 | 32 – 40 | 33 – 41 | 32 – 40 | 34 – 42 | 34 – 42 | 35 – 43 | 34 – 43 |
| 29 – 36 | 29 – 36 | 29 – 37 | 30 – 37 | 29 – 37 | **31 – 38** | 30 – 38 | 31 – 38 | 32 – 39 | 31 – 39 | **32 – 40** | 32 – 40 | 33 – 41 | 33 – 41 |
| 28 – 35 | 28 – 35 | 29 – 36 | 29 – 36 | 28 – 36 | **30 – 37** | 29 – 36 | 29 – 37 | 30 – 38 | 30 – 37 | **31 – 38** | 31 – 38 | 31 – 39 | 31 – 39 |
| 32 – 40 | 33 – 41 | 33 – 42 | 34 – 42 | 33 – 42 | **35 – 44** | 35 – 43 | 35 – 44 | 36 – 45 | 36 – 45 | 37 – 47 | 37 – 46 | 38 – 48 | 38 – 47 |
| 31 – 39 | 32 – 39 | 32 – 40 | 33 – 40 | 32 – 40 | **34 – 42** | 33 – 41 | 34 – 42 | 35 – 43 | 34 – 43 | 35 – 44 | 35 – 44 | 36 – 45 | 36 – 45 |
| 30 – 38 | 30 – 38 | 31 – 39 | 31 – 39 | 31 – 39 | **32 – 40** | 32 – 39 | 32 – 40 | 33 – 41 | 32 – 41 | 34 – 42 | 33 – 42 | 34 – 43 | 34 – 42 |
| 35 – 44 | 36 – 44 | 37 – 46 | 37 – 46 | 37 – 46 | **39 – 48** | 38 – 48 | 39 – 49 | 40 – 50 | 40 – 49 | 41 – 52 | 41 – 52 | 43 – 53 | 43 – 53 |
| 34 – 42 | 34 – 43 | 35 – 44 | 35 – 44 | 35 – 44 | **37 – 46** | 36 – 45 | 37 – 46 | 38 – 47 | 38 – 47 | 39 – 49 | 39 – 49 | 40 – 50 | 40 – 50 |
| 33 – 41 | 33 – 41 | 33 – 42 | 34 – 42 | 34 – 42 | **35 – 44** | 34 – 43 | 35 – 44 | 36 – 45 | 35 – 44 | 37 – 46 | 36 – 45 | 37 – 47 | 37 – 46 |
| 38 – 47 | 38 – 48 | 39 – 49 | 40 – 50 | 39 – 49 | 41 – 52 | 41 – 51 | 42 – 52 | 43 – 53 | 42 – 53 | 44 – 55 | 45 – 56 | **46 – 57** | 46 – 57 |
| 22 – 44[o] | 27 – 62[n] | ... | 31 – 57[o] | 26 – 49[q] | 49 – 85[r] | 24 – 46[n] | < 76[n] | ... | ... | 42 – 69[t] | 33 – 62[n] | 34 – 45[u] | < 59[n] |
| 29[o] | 41[n] | ... | 40[o] | 36[q] | 64[r] | 32[n] | 12[n] | ... | ... | 54[t] | 44[n] | 39[u] | 17[n] |

[o]Data-weighted $W_l$ (Gundersen et al. 1995), with 15% calibration uncertainty. [p]SK95C$n$ and SK95R$n$ are the SK95 Q band NCP cap and ring $n$-point chops (Netterfield et al. 1996). [q](Netterfield 1995), with 15% calibration uncertainty. [r]PYTHG,I are the PYTHON additive large- and small-chop $W_l$ (Alvarez et al. 1996). Calibration uncertainty is 20%. [s](Halpern et al. 1993). [t]PYTHL,S are the PYTHON subtractive large- and small-chop $W_l$ (Platt et al. 1995), with three years of data for L, and one year for S. Calibration uncertainty is 20%. [u](de Bernardis et al. 1994), fiducial CDM, with 5% calibration uncertainty.

| (108, 94) | (114, 127) | (115, 115) | (120, 115) | (125, 123) | (133, 145) | (134, 134) | (135, 117) | (139, 134) | (139, 150) | (142, 155) | (144, 157) | (152, 153) | (157, 140) | (157, 154) |
|---|---|---|---|---|---|---|---|---|---|---|---|---|---|---|
| (76, 115) | (70, 196) | (95, 136) | (95, 136) | (84, 168) | (80, 224) | (115, 155) | (97, 141) | (115, 156) | (83, 232) | (85, 240) | (86, 243) | (134, 175) | (119, 165) | (134, 176) |
| 0.810 | 1.41 | 0.620 | 0.665 | 0.895 | 1.51 | 0.542 | 0.755 | 0.586 | 1.55 | 1.49 | 1.51 | 0.481 | 0.667 | 0.524 |
| SK95C5[p] | MAX4 6,9[v] | SK94K6[n] | SK94Q6[n] | PYTHI[r] | MAX4 3.5[w] | SK94K7[n] | SK95C6[p] | SK94Q7[n] | MAX5[x] | MAX3 G[y] | MAX3 M[z] | SK94K8[p] | SK95C7[p] | SK94Q8[p] |
| ($\mu$K) | ($\mu$K) | ($\mu$K) | ($\mu$K) | ($\mu$K) | ($\mu$K) | ($\mu$K) | ($\mu$K) | ($\mu$K) | ($\mu$K) | ($\mu$K) | ($\mu$K) | ($\mu$K) | ($\mu$K) | ($\mu$K) |
| (28) | (29) | (30) | (31) | (32) | (33) | (34) | (35) | (36) | (37) | (38) | (39) | (40) | (41) | (42) |
| 24 – 30 | 25 – 31 | 24 – 30 | 24 – 30 | 25 – 31 | **26 – 32** | 25 – 31 | 25 – 32 | 25 – 32 | 26 – 33 | **26 – 33** | 26 – 33 | 26 – 33 | 27 – 33 | 27 – 33 |
| 30 – 37 | 31 – 39 | 30 – 38 | 31 – 39 | 32 – 40 | **33 – 41** | 32 – 40 | 33 – 41 | 33 – 41 | 34 – 42 | **34 – 42** | 34 – 43 | 34 – 42 | 35 – 43 | 34 – 43 |
| 29 – 36 | 30 – 37 | 29 – 36 | 30 – 37 | 30 – 38 | **31 – 39** | 31 – 38 | 31 – 39 | 31 – 39 | 32 – 40 | **32 – 40** | 32 – 40 | 32 – 40 | 33 – 41 | 33 – 41 |
| 28 – 34 | 28 – 35 | 28 – 35 | 28 – 35 | 29 – 36 | **30 – 37** | 29 – 36 | 29 – 37 | 29 – 37 | 30 – 38 | **31 – 38** | 31 – 38 | 30 – 38 | 31 – 39 | 31 – 38 |
| 35 – 44 | 37 – 46 | 36 – 45 | 37 – 46 | 38 – 47 | 39 – 49 | 38 – 48 | 39 – 48 | 39 – 48 | 40 – 50 | **40 – 50** | **41 – 51** | 41 – 51 | 41 – 52 | 41 – 51 |
| 34 – 42 | 35 – 43 | 34 – 42 | 35 – 43 | 36 – 44 | 37 – 46 | 36 – 45 | 36 – 46 | 37 – 46 | 38 – 47 | **38 – 47** | **38 – 48** | 38 – 47 | 39 – 48 | 39 – 48 |
| 32 – 40 | 33 – 41 | 32 – 40 | 33 – 41 | 34 – 42 | 35 – 43 | 34 – 42 | 34 – 43 | 34 – 43 | 35 – 44 | **35 – 44** | 36 – 44 | 35 – 44 | 36 – 45 | 36 – 45 |
| 39 – 49 | 41 – 51 | 40 – 50 | 41 – 51 | 42 – 52 | 44 – 54 | 43 – 53 | 43 – 54 | 43 – 54 | 44 – 55 | **45 – 56** | **45 – 56** | 45 – 57 | 46 – 58 | 46 – 58 |
| 37 – 46 | 38 – 48 | 38 – 47 | 38 – 48 | 39 – 49 | 41 – 51 | 40 – 50 | 40 – 50 | 41 – 51 | 41 – 52 | **42 – 52** | **42 – 52** | 42 – 53 | 43 – 54 | 43 – 54 |
| 35 – 43 | 36 – 45 | 35 – 44 | 36 – 44 | 37 – 46 | 38 – 47 | 37 – 46 | 37 – 47 | 37 – 47 | 38 – 48 | **39 – 48** | **39 – 49** | 39 – 49 | 40 – 50 | 40 – 49 |
| 44 – 55 | 46 – 57 | 45 – 56 | 46 – 57 | 47 – 59 | 49 – 61 | 48 – 60 | 48 – 60 | 49 – 61 | 49 – 62 | 50 – 62 | **50 – 63** | 51 – 64 | 52 – 65 | 52 – 65 |
| 41 – 51 | 43 – 53 | 42 – 52 | 42 – 53 | 44 – 55 | 45 – 56 | 45 – 56 | 45 – 56 | 45 – 57 | 46 – 57 | **46 – 58** | **47 – 58** | 47 – 59 | 48 – 60 | 48 – 60 |
| 38 – 48 | 40 – 49 | 39 – 48 | 39 – 49 | 41 – 51 | 42 – 52 | 41 – 51 | 41 – 52 | 42 – 52 | 42 – 53 | **43 – 53** | **43 – 54** | 44 – 54 | 44 – 55 | 44 – 55 |
| 47 – 58 | 49 – 61 | 49 – 61 | 50 – 62 | 50 – 63 | 51 – 63 | 53 – 66 | 52 – 65 | 54 – 67 | 51 – 64 | 52 – 64 | **52 – 65** | 57 – 71 | 56 – 70 | 58 – 72 |
| ... | 29 – 68[v] | 22 – 50[n] | < 63[n] | ... | 42 – 87[w] | < 76[n] | ... | < 120[n] | 25 – 71[x] | 60 – 96[y] | 16 – 36[z] | < 130[p] | ... | < 120[p] |
| ... | ... | 33[n] | 25[n] | ... | ... | 35[n] | ... | 67[n] | ... | 74[y] | 23[z] | 66[p] | ... | 64[p] |

[v]MAX4 6 and 9 cm$^{-1}$ $W_l$ (Devlin et al. 1994; Clapp et al. 1994; as reanalyzed by Tanaka et al. 1995, hereafter T95; individual channel observational numbers are courtesy of S. Tanaka, private communication 1995). The observational range is a composite of 3 sets of 6 and 9 cm$^{-1}$ measurements — $42^{+18}_{-13}\mu$K, $54^{+25}_{-17}\mu$K (GUM); $(15, < 72)\mu$K, $42^{+26}_{-18}\mu$K (SH); $51^{+29}_{-21}\mu$K, $41^{+27}_{-19}\mu$K (ID) — estimated by discarding the two largest upper limits and the two smallest lower limits (among the five detections). Observational error bars account for calibration uncertainty of 10%. [w]MAX4 3.5 cm$^{-1}$ $W_l$. See previous footnote. The observational range is a composite of 3 sets of 3.5 cm$^{-1}$ measurements — $79^{+29}_{-23}\mu$K (GUM); $61^{+26}_{-19}\mu$K (SH); $56^{+26}_{-18}\mu$K (ID) — estimated by discarding the largest upper limit and the smallest lower limit. [x]MAX5 (T95). The observational range is a composite of two measurements — $33^{+11}_{-8}\mu$K, $52^{+19}_{-11}\mu$K (HR5127, PH; T95) — estimated using the largest upper limit and the smallest lower limit. Calibration uncertainty is 10%. A third MAX5 scan is under analysis (Lim et al. 1996). [y]MAX3 GUM (Gundersen et al. 1993, as recomputed by J. Gundersen, private communication 1995). All MAX GUM $W_l$ are for a single scan and ignore sky rotation. Calibration uncertainty is 10%. [z]MAX3 MUP (Meinhold et al. 1993, as recomputed by J. Gundersen, private communication 1995). Note that dust has been subtracted.

| (159, 151) | (170, 171) | (170, 173) | (176, 173) | (177, 160) | (178, 183) | (197, 180) | (217, 199) | (234, 237) | (237, 220) | (257, 241) | (263, 270) | (277, 262) | (286, 302) | (297, 283) |
|---|---|---|---|---|---|---|---|---|---|---|---|---|---|---|
| (83, 234) | (115, 236) | (153, 194) | (153, 195) | (140, 184) | (133, 239) | (159, 204) | (178, 222) | (182, 301) | (199, 244) | (221, 265) | (181, 375) | (241, 286) | (247, 365) | (263, 307) |
| 1.45 | 1.13 | 0.432 | 0.474 | 0.594 | 0.684 | 0.543 | 0.496 | 0.852 | 0.461 | 0.425 | 1.14 | 0.398 | 0.702 | 0.371 |
| MSAM2[aa] | SK95R3[p] | SK94K9[p] | SK94Q9[p] | SK95C8[p] | PYTHS[t] | SK95C9[p] | SK95C10[p] | SK95R4[p] | SK95C11[p] | SK95C12[p] | MSAM3[bb] | SK95C13[p] | SK95R5[p] | SK95C14[p] |
| ($\mu$K) | ($\mu$K) | ($\mu$K) | ($\mu$K) | ($\mu$K) | ($\mu$K) | ($\mu$K) | ($\mu$K) | ($\mu$K) | ($\mu$K) | ($\mu$K) | ($\mu$K) | ($\mu$K) | ($\mu$K) | ($\mu$K) |
| (43) | (44) | (45) | (46) | (47) | (48) | (49) | (50) | (51) | (52) | (53) | (54) | (55) | (56) | (57) |
| 28 – 34 | 28 – 34 | 27 – 34 | **28 – 34** | 28 – 35 | **28 – 35** | 29 – 36 | 30 – 37 | 31 – 39 | 31 – 39 | 32 – 40 | 33 – 41 | 33 – 42 | 34 – 43 | 35 – 43 |
| 36 – 44 | 36 – 45 | 36 – 44 | **36 – 45** | 37 – 46 | 37 – 46 | 38 – 48 | 40 – 50 | 42 – 53 | 42 – 53 | 44 – 55 | 45 – 56 | 46 – 57 | 47 – 59 | 48 – 60 |
| 34 – 42 | 34 – 43 | 34 – 42 | **34 – 43** | 35 – 43 | **35 – 43** | 36 – 45 | 38 – 47 | 40 – 49 | 40 – 49 | 41 – 51 | 42 – 52 | 43 – 54 | 44 – 55 | 45 – 56 |
| 32 – 40 | 32 – 40 | 32 – 40 | **32 – 40** | 32 – 40 | **33 – 41** | 34 – 42 | 35 – 44 | 37 – 46 | 37 – 46 | 38 – 48 | 39 – 48 | 40 – 50 | 41 – 51 | 41 – 51 |
| 42 – 52 | 43 – 54 | 43 – 53 | **44 – 54** | 44 – 55 | 44 – 55 | 46 – 58 | 49 – 61 | 51 – 64 | 51 – 64 | 54 – 67 | **53 – 66** | 56 – 70 | 57 – 70 | 58 – 72 |
| 39 – 49 | 41 – 51 | 40 – 50 | **41 – 51** | 41 – 51 | 41 – 52 | 43 – 54 | 45 – 57 | 47 – 59 | 48 – 59 | 50 – 62 | 49 – 61 | 52 – 65 | 52 – 65 | 54 – 67 |
| 37 – 46 | 38 – 47 | 37 – 46 | **38 – 47** | 38 – 47 | 38 – 48 | 40 – 50 | 42 – 52 | 44 – 54 | 44 – 55 | 46 – 57 | 45 – 56 | 48 – 59 | 48 – 60 | 49 – 62 |
| **46 – 57** | 48 – 60 | 48 – 60 | **49 – 61** | 49 – 61 | 50 – 62 | 52 – 65 | 55 – 68 | 57 – 71 | 57 – 72 | 60 – 75 | **58 – 72** | 62 – 78 | 61 – 77 | 64 – 80 |
| 43 – 53 | 45 – 56 | 45 – 56 | **45 – 57** | 46 – 57 | 46 – 57 | 48 – 60 | 51 – 63 | 52 – 65 | 53 – 66 | 55 – 69 | **53 – 66** | 57 – 71 | 56 – 70 | 59 – 73 |
| 40 – 49 | 41 – 52 | 41 – 51 | **42 – 52** | 42 – 52 | 42 – 53 | 44 – 55 | 46 – 58 | 48 – 60 | 48 – 60 | 51 – 63 | 48 – 60 | 52 – 65 | 51 – 64 | 54 – 67 |
| **51 – 63** | 54 – 68 | 55 – 68 | **56 – 69** | 56 – 69 | 56 – 70 | 59 – 73 | 62 – 77 | 63 – 79 | 65 – 81 | 67 – 84 | **63 – 79** | 69 – 86 | 67 – 84 | 70 – 88 |
| **47 – 59** | 50 – 62 | 50 – 63 | **51 – 64** | 51 – 64 | 52 – 64 | 54 – 67 | 57 – 71 | 58 – 72 | 59 – 74 | 61 – 76 | **57 – 72** | 63 – 79 | 61 – 76 | 64 – 80 |
| 43 – 54 | 46 – 57 | 46 – 58 | **47 – 59** | 47 – 58 | 47 – 59 | 49 – 62 | 52 – 65 | 53 – 66 | 54 – 67 | 56 – 70 | 52 – 65 | 57 – 71 | 55 – 69 | 58 – 72 |
| **51 – 64** | 56 – 69 | 60 – 75 | **60 – 75** | 59 – 73 | 58 – 73 | 61 – 76 | 62 – 77 | 59 – 73 | 62 – 77 | 61 – 76 | **56 – 70** | 58 – 73 | 54 – 67 | 55 – 69 |
| 24 – 44[aa] | ... | < 150[p] | 90 – 200[p] | ... | 46 – 73[t] | ... | ... | ... | ... | ... | 30 – 52[bb] | ... | ... | ... |
| 33[aa] | ... | 5.8[p] | 140[p] | ... | 59[t] | ... | ... | ... | ... | ... | 40[bb] | ... | ... | ... |

[aa]Second year MSAM 2-beam $W_l$ (Cheng et al. 1995, hereafter C95). Unless given in C95, MSAM central values (cv) are the geometric mean of the upper and lower $2\sigma$ values, and the $1\sigma$ values are = $\text{cv}(2\sigma/\text{cv})^{1/1.7}$ (D. Cottingham, private communication 1995), with 10% calibration uncertainty. The first year measurement gives $53^{+15}_{-12}\mu$K (C95). [bb] Second year MSAM 3-beam $W_l$ (C95). See previous footnote. The first year measurement gives $53^{+12}_{-11}\mu$K (C95).

| SK95C15[p] | WDH1[cc] | SK95R6[p] | SK95C16[p] | SK95C17[p] | SK95C18[p] | SK95C19[p] | WDI[dd] | WDH2[cc] | OVROL[ee] | WDH3[cc] | OVROS[ee] | SUZIL2[ff] | SUZIS2[ff] | SUZIS3[ff] |
|---|---|---|---|---|---|---|---|---|---|---|---|---|---|---|
| (316, 302) | (329, 382) | (332, 369) | (333, 322) | (357, 346) | (382, 371) | (404, 394) | (477, 539) | (559, 608) | (598, 537) | (759, 804) | (2020, 1700) | (2360, 2410) | (3250, 3670) | (4010, 4210) |
| (282, 326) | (213, 570) | (315, 431) | (301, 345) | (325, 369) | (352, 393) | (374, 416) | (297, 825) | (412, 815) | (361, 754) | (591, 1030) | (1140, 2390) | (1340, 3680) | (2020, 5600) | (2830, 5770) |
| 0.352 | 1.56 | 0.585 | 0.334 | 0.312 | 0.288 | 0.274 | 1.18 | 0.781 | 1.41 | 0.415 | 1.50 | 1.73 | 1.27 | 0.933 |
| ($\mu$K) | ($\mu$K) | ($\mu$K) | ($\mu$K) | ($\mu$K) | ($\mu$K) | ($\mu$K) | ($\mu$K) | ($\mu$K) | ($\mu$K) | ($\mu$K) | ($\mu$K) | ($\mu$K) | ($\mu$K) | ($\mu$K) |
| (58) | (59) | (60) | (61) | (62) | (63) | (64) | (65) | (66) | (67) | (68) | (69) | (70) | (71) | (72) |
| 36 − 44 | 36 − 45 | 37 − 46 | 36 − 46 | 38 − 47 | 39 − 49 | 40 − 50 | 42 − 52 | 45 − 57 | 43 − 54 | 48 − 60 | 36 − 44 | 34 − 42 | 29 − 36 | 24 − 30 |
| 49 − 62 | 49 − 60 | 51 − 63 | 51 − 63 | 53 − 66 | 55 − 68 | 56 − 70 | **51 − 63** | 55 − 68 | 53 − 66 | 50 − 63 | 36 − 45 | 34 − 43 | 28 − 35 | 21 − 26 |
| 46 − 57 | 45 − 56 | 47 − 59 | 47 − 59 | 49 − 61 | 51 − 63 | 52 − 65 | 47 − 58 | 50 − 62 | 49 − 61 | 46 − 57 | 34 − 43 | 32 − 40 | 26 − 32 | 19 − 24 |
| 43 − 53 | 42 − 52 | 44 − 55 | 44 − 55 | 45 − 57 | 47 − 59 | 48 − 60 | 43 − 54 | 45 − 57 | 45 − 56 | 43 − 53 | 32 − 39 | 29 − 36 | 23 − 29 | 16 − 20 |
| 60 − 75 | 55 − 69 | 60 − 75 | 61 − 77 | 63 − 79 | 64 − 80 | 65 − 81 | **54 − 68** | 56 − 70 | 56 − 70 | 49 − 62 | 37 − 46 | 34 − 43 | 27 − 34 | 19 − 23 |
| 55 − 69 | 50 − 63 | 55 − 68 | 57 − 71 | 58 − 72 | 59 − 74 | 59 − 74 | 50 − 62 | 51 − 63 | 51 − 63 | 46 − 58 | 34 − 43 | 32 − 39 | 25 − 31 | 16 − 20 |
| 51 − 63 | 46 − 58 | 50 − 63 | 52 − 65 | 53 − 66 | 54 − 67 | 54 − 68 | 46 − 57 | 47 − 58 | 47 − 58 | 44 − 55 | 31 − 38 | 28 − 35 | 21 − 27 | 13 − 16 |
| 65 − 82 | 57 − 71 | 62 − 78 | 66 − 83 | 67 − 83 | 67 − 83 | 66 − 82 | **54 − 68** | 54 − 67 | 54 − 68 | 48 − 60 | 35 − 44 | 33 − 41 | 25 − 31 | 16 − 20 |
| 60 − 74 | 52 − 64 | 56 − 70 | 60 − 75 | 60 − 75 | 60 − 75 | 59 − 73 | 50 − 63 | 50 − 62 | 50 − 62 | 46 − 58 | 33 − 41 | 30 − 37 | 23 − 28 | 14 − 17 |
| 54 − 68 | 47 − 59 | 51 − 64 | 55 − 68 | 55 − 69 | 54 − 68 | 53 − 66 | 47 − 58 | 47 − 58 | 46 − 58 | 44 − 55 | 29 − 36 | 26 − 33 | 19 − 24 | 11 − 13 |
| 71 − 89 | 60 − 75 | 65 − 81 | 71 − 89 | 71 − 88 | 69 − 86 | 67 − 83 | **57 − 71** | 56 − 69 | 56 − 70 | 51 − 63 | 35 − 44 | 33 − 41 | 25 − 31 | 15 − 19 |
| 64 − 80 | 55 − 68 | 58 − 73 | 64 − 80 | 63 − 78 | 61 − 76 | 58 − 73 | **53 − 66** | 52 − 65 | 52 − 65 | 49 − 61 | 32 − 40 | 30 − 37 | 22 − 28 | 12 − 16 |
| 58 − 73 | 50 − 63 | 53 − 66 | 58 − 72 | 57 − 71 | 55 − 68 | 52 − 65 | 49 − 61 | 49 − 61 | 48 − 60 | 46 − 57 | 28 − 35 | 26 − 32 | 19 − 24 | 9.5 − 12 |
| 52 − 65 | 53 − 66 | 48 − 60 | 49 − 61 | 46 − 58 | 45 − 56 | 46 − 57 | 50 − 63 | 49 − 61 | 48 − 60 | 46 − 57 | 24 − 30 | 23 − 29 | 17 − 21 | 7.1 − 8.8 |
| ... | 43 − 120[cc] | ... | ... | ... | ... | ... | < 50[dd] | < 88[cc] | ... | < 160[cc] | ... | ... | ... | ... |
| ... | 77[cc] | ... | ... | ... | ... | ... | 0[dd] | 30[cc] | ... | 60[cc] | ... | ... | ... | ... |

[cc]WDH$n$ are the WD $n^{\rm th}$-harmonic $W_l$ (Griffin et al. 1995), with calibration uncertainty of 30%. [dd](Tucker et al. 1993). Calibration uncertainty is accounted for by multiplying the $2\sigma$ upper limit by $\sqrt{1.3}$ (J. Peterson, private communication 1995). [ee](Myers, Leitch, & Readhead 1996). [ff]SUZIL2, S2, and S3 are the SuZIE large 2-beam, and small 2- and 3-beam $W_l$ (Church et al. 1996; Ganga et al. 1996).

# CMB ANISOTROPY IN COBE-DMR-NORMALIZED OPEN CDM COSMOGONY

Bharat Ratra[1], Anthony J. Banday[2], Krzysztof M. Górski[3,4], and Naoshi Sugiyama[5]

[1]*Joseph Henry Laboratories, Princeton University, Princeton, NJ 08544*

[2]*Hughes STX Corporation, 4400 Forbes Bldg, Lanham, MD 20706*

[3]*Universities Space Research Association, NASA/GSFC, Code 685, Greenbelt, MD 20771*

[4]*Warsaw University Observatory, Aleje Ujazdowskie 4, 00-478 Warszawa, Poland*

[5]*Department of Physics and Research Center for the Early Universe,*

*University of Tokyo, Tokyo 113, Japan*

## ABSTRACT

We compute the CMB anisotropy in an open inflation CDM model which is normalized to the two-year DMR sky map. Presently available CMB data is consistent with $\Omega_0 \sim 0.3 - 0.4$, but does not strongly disfavour either $\Omega_0 \sim 0.1$ or $\Omega_0 \sim 1$. Accumulating CMB data should strengthen the observational discriminative power, however, a better understanding of systematic effects will be needed before it will be possible to draw robust conclusions about model viability.

*Subject headings:* cosmic microwave background — cosmology: observations — large-scale structure of the universe — galaxies: formation

# 1. INTRODUCTION

Observational evidence indicates that a low-$\Omega_0$ cosmogony might not be an unreasonable model of the universe. Of interest is a low-density CDM model with open spatial sections and no cosmological constant, with an initial epoch of inflation (Ratra & Peebles 1995, hereafter RP)[7]. Normalized to the two-year DMR sky map (in galactic coordinates) (Górski et al. 1995, hereafter GRSB), this model is mostly consistent with large-scale structure observations when $\Omega_0 \sim 0.3 - 0.4$ (GRSB; Liddle et al. 1995, hereafter LLRV). Here we examine the compatability of the primary[8] CMB anisotropy predictions of this model with what has been measured on sub-DMR scales.

# 2. SUMMARY OF COMPUTATION

The parameters characterizing models 1 – 13 (Table 1, col. [2]) are chosen to be roughly consistent with small-scale estimates of $\Omega_0$, measurements of $h$ and $t_0$, and nucleosynthesis bounds on $\Omega_B h^2$. (Model 14 is the usual fiducial CDM case.) For each $\Omega_0$, the range of $h$ and $\Omega_B h^2$ was chosen to widen the spread in the CMB anisotropy predictions. Effects of early reionization, tilt, and gravity waves are ignored, as they are unlikely to be significant in viable open models.

The CMB anisotropy computation (Sugiyama 1995, hereafter S95) improves upon that used previously by accounting for polarization and using a better approximation to recombination. The DMR-scale $C_l$ are unaffected, and, since the $C_l$ here are almost entirely governed by $\Omega_0$, we use the GRSB (quadrupole-excluded) normalization. With $\delta T/T = \sum_{l,m} a_{lm} Y_{lm}$, the rms temperature anisotropy, seen through a window $W_l$, is $(\delta T/T)_{\rm rms} = (\sum_{l=2}^{\infty} (2l+1) C_l W_l/(4\pi))^{1/2}$, where $C_l = \langle |a_{lm}|^2 \rangle$. Defining $I(W_l) = \sum_{l=2}^{\infty}(l+0.5)W_l/[l(l+1)]$, the bandtemperature, which is relatively insensitive to the detailed shape of $W_l$, is $\delta T_l = \delta T_{\rm rms}/\sqrt{I(W_l)}$, and the effective multipole $l_{\rm e} = I(lW_l)/I(W_l)$ (Bond 1995). The computation of the large-scale structure parameters (cols. [7–9]) (S95) improves upon what was used for GRSB, and changes the predictions by $1-3\%$. Here we also follow Stompor, Górski, & Banday (1995) and account for systematic uncertainties in the DMR normalization, by assigning total ($1\sigma$) error bars of $\pm 11\%$ to the estimated $Q_{\rm rms-PS}^{\rm (no-Q)}$. This accounts for the difference between the galactic and ecliptic coordinate maps, the effect of including or excluding the quadrupole in the analysis, the effects of varying $h$ and $\Omega_B$ on the DMR-scale $C_l$ (GRSB normalized at fixed $h$ and $\Omega_B$), and the numerical uncertainty in the $C_l$ computation.

---

[7]The perturbation power spectrum, derived in RP, and valid for inflation-epoch potentials that satisfy the slow-roll constraint, has also been derived by Lyth & Woszczyna (1995, hereafter LW) and White & Bunn (1995), under the assumption of further approximations. This model is an approximation of that where an open-inflation bubble nucleates in an inflating, spatially-flat, de Sitter spacetime. Accounting for the nucleation process (Bucher & Turok 1995, hereafter BT; Yamamoto, Sasaki, & Tanaka 1995, hereafter YST) shows that, because of the spatial curvature 'cutoff' in an open universe (RP), predictions of observationally viable nucleation models agree with those derived using the simple (RP) spectrum (YST; BT). There is also the extreme option (inconsistent with standard quantum mechanics in an open universe) of allowing non-square-integrable basis functions (LW; YST).

[8]For this class of models, secondary CMB anisotropies are likely to be insignificant on all but the smallest scales we consider here.

We used analytic expressions to construct $W_l$, except for MSAM (the $W_l$ are nongaussian, were constructed by A. Kosowsky, and are courtesy of L. Page) and for SK94 and SK95 (the differencing is done in software and the $W_l$ are courtesy of B. Netterfield). The FIRS $W_l$ does not account for beam smearing, and the SK93 one only approximately does (it is off by $1-2\%$, and this is the largest difference we are aware of). The main uncertainty in the $W_l$ is that due to $W_l$-parameter measurement errors. While it would be preferable to fold these into the theoretical predictions, in most cases the calibration uncertainty is large ($\sim 15\%$), and, since we account for this in the observational error bars, we ignore smaller $W_l$-parameter measurement uncertainties. The first line of the Table gives $l_{\rm e}$, and $l_{\rm m}$, the multipole where $W_l$ peaks. The second gives $l_{e-0.5}$, the two multipoles where $W_{l_{e-0.5}} = e^{-0.5} W_{l_m}$ (except for FIRS). The DMR-normalized $1\sigma$ range of the bandtemperature predictions for models $1-14$ are tabulated for the experimental $W_l$ in columns (10) – (72) of the Table.

The various experimental groups have analyzed their data assuming that the CMB anisotropy signal can be approximated by that corresponding to either a gaussian autocorrelation function, a flat bandpower spectrum, or the CMB spectrum computed in the fiducial CDM model, and quote a central value (and an error range) for the amplitude of the assumed spectrum that best reproduces their data. When necessary, we have converted these quoted central values and error ranges to bandtemperature, $\delta T_l$, central values and error ranges. In the table, these CMB data have $1\sigma$ error bars if there is a $2\sigma$ detection away from 0 (otherwise, they are $2\sigma$ upper limits). These error bars account for the size of the observational sample, as well as the uncertainty introduced by there being only one observable universe. Although not strictly proper, absolute calibration uncertainty has been added in quadrature to the error bars (twice the calibration uncertainty for those $2\sigma$ upper limits where the likelihood peaks away from 0). The last line of the table gives the central value, and the penultimate line the range, of the observational data. Given that a number of the experiments have not yet confirmed their initial discovery with a follow-up run, and that for a number of cases it is not yet known whether or not the observed sky anisotropy signals are purely CMB anisotropy signals (i.e., non-CMB foreground contamination could be a problem), due caution must be exercised when comparing observational results to theoretical predictions. We also emphasize that different groups use different statistical techniques to analyze the observational data (e.g., maximum likelihood, likelihood ratio), and also use different statistical prescriptions to define their central values (e.g., mode, median) as well as their error ranges (e.g., 68.3% highest posterior density region, 16% and 84% bayesian limits), so due caution must also be exercised when comparing different observational results tabulated here.

## 3. DISCUSSION

The boldface entries in cols. (4), (8), and (9) of the table are disfavoured by what are now thought to be reasonable large-scale structure estimates. The parameters can be further constrained by using other data. Estimates of cluster baryon fractions (White et al. 1993) are difficult to reconcile with models 14, 13, and 10, and put pressure on 7, 12, and 9, and cluster abundances (LLRV) disfavour the $\delta M/M(8h^{-1}\mathrm{Mpc})$ values of models $1-6$ and 14, and put pressure on 7 (which can again be eased by slightly adjusting parameters).

CMB experiments with multiple windows have nonzero correlation between windows, which must be accounted for when comparing a model to data. While the relative calibration uncertainty between windows of a multiple window experiment is small, we wish to simplify the problem by fixing the normalization to that from the DMR and use the smaller-scale experiments to only constrain the shape, so we have accounted for the absolute calibration uncertainty in the error bars. Whilst it is clear that better control of systematic uncertainties will be needed before it will be possible to draw robust conclusions about the viability of models, it is encouraging that data (close in $l$-space) from different experiments, and for different parts of the sky, is not completely inconsistent.

It is also encouraging that after normalizing these models to the DMR, the smaller-scale CMB constraints are not completely inconsistent with those from the classical cosmological tests, standard nucleosynthesis, and large-scale structure data. We emphasize, however, that present CMB anisotropy observational data does not strongly discriminate between models. As indicated by the boldface entries in the table, an $\Omega_0 \lesssim 0.1$ model is not consistent with cosmological data. It is interesting that $\Omega_0 \sim 0.3 - 0.4$ is mostly consistent with the data, and that $\Omega_0 \sim 1$ is not favoured. If future data do not rule out $\Omega_0 \sim 0.3 - 0.4$, the puzzle for the model will be whether there really is baryonic matter that does not take part in nucleosynthesis, or whether the required CDM is nonbaryonic.

We acknowledge the very generous assistance of D. Alvarez, D. Cottingham, P. de Bernardis, M. Dragovan, S. Hancock, M. Lim, S. Platt, L. Piccirillo, T. Readhead, J. Ruhl, G. Tucker, and especially M. Devlin, K. Ganga, G. Griffin, J. Gundersen, B. Netterfield, S. Tanaka, as well as L. Page. This work was supported in part by NSF grant PHY89-21378, and by the NASA Office of Space Sciences.

FIGURE CAPTIONS

Fig. 1.– CMB anisotropy bandtemperature predictions for models 1 (highest at $l \sim 2000$), 9, and 14 (highest at $l \sim 200$). Continuous lines are what would be seen by a series of ideal, Kronecker window ($W_l \propto \delta_{l,l_e}$), experiments, for the models normalized to the central values of the DMR normalization of GRSB. Open squares (placed at the appropriate $l_e$, and only for models 1 and 14) are the predictions for the $W_l$ of the table, with horizontal lines terminating at $l_{e-0.5}$, and with vertical, correlated, $1\sigma$ error bars from the DMR normalization of GRSB. Note that for wider $W_l$, especially those at larger $l$, the theoretical predictions for the real $W_l$ significantly deviate from what would be seen by ideal Kronecker $W_l$. While near-future CMB observations should be able to distinguish between models 1 and 14, it must be kept in mind that both these models are difficult to reconcile with large-scale structure data, and that models consistent with large-scale structure constraints lie between these two extreme cases (for most of $l$-space). As $l_e$ is not particularly physical, it is more meaningful to directly compare the model predictions to the data points shown in Fig. 2.

Fig. 2.– CMB anisotropy bandtemperature observations (placed at the appropriate $l_e$) for all individual $W_l$ with data in the table (including those in the footnotes), except SK94Q9 which is just off scale. Open squares are detections that are at least $2\sigma$ away from 0, and triangles are $2\sigma$ upper limits (they are placed at the $2\sigma$ upper limit, not at the peak of the likelihood). Vertical $1\sigma$ error bars also account for absolute calibration uncertainty. The continuous lines are the models of Fig. 1.

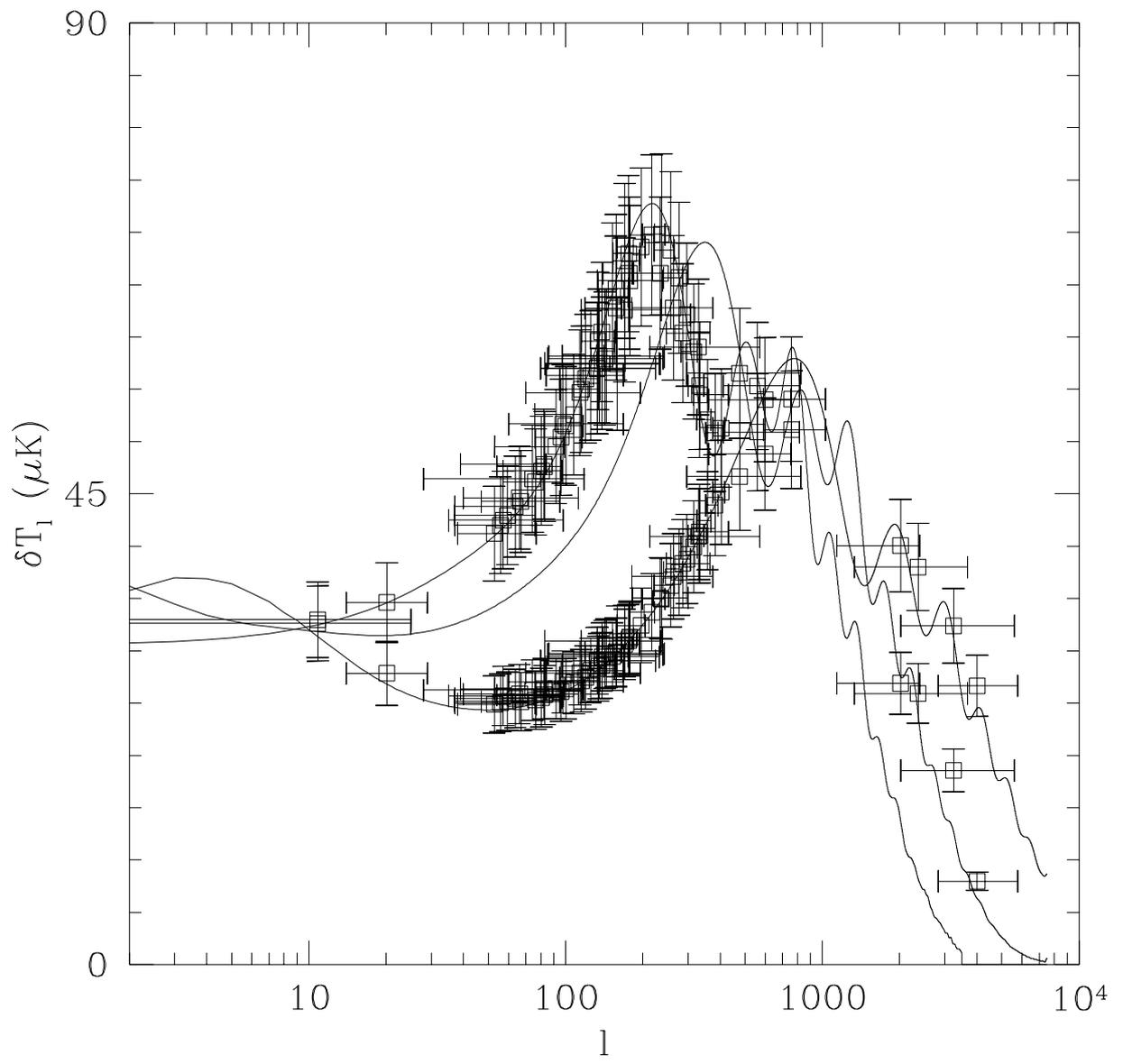

Figure 1

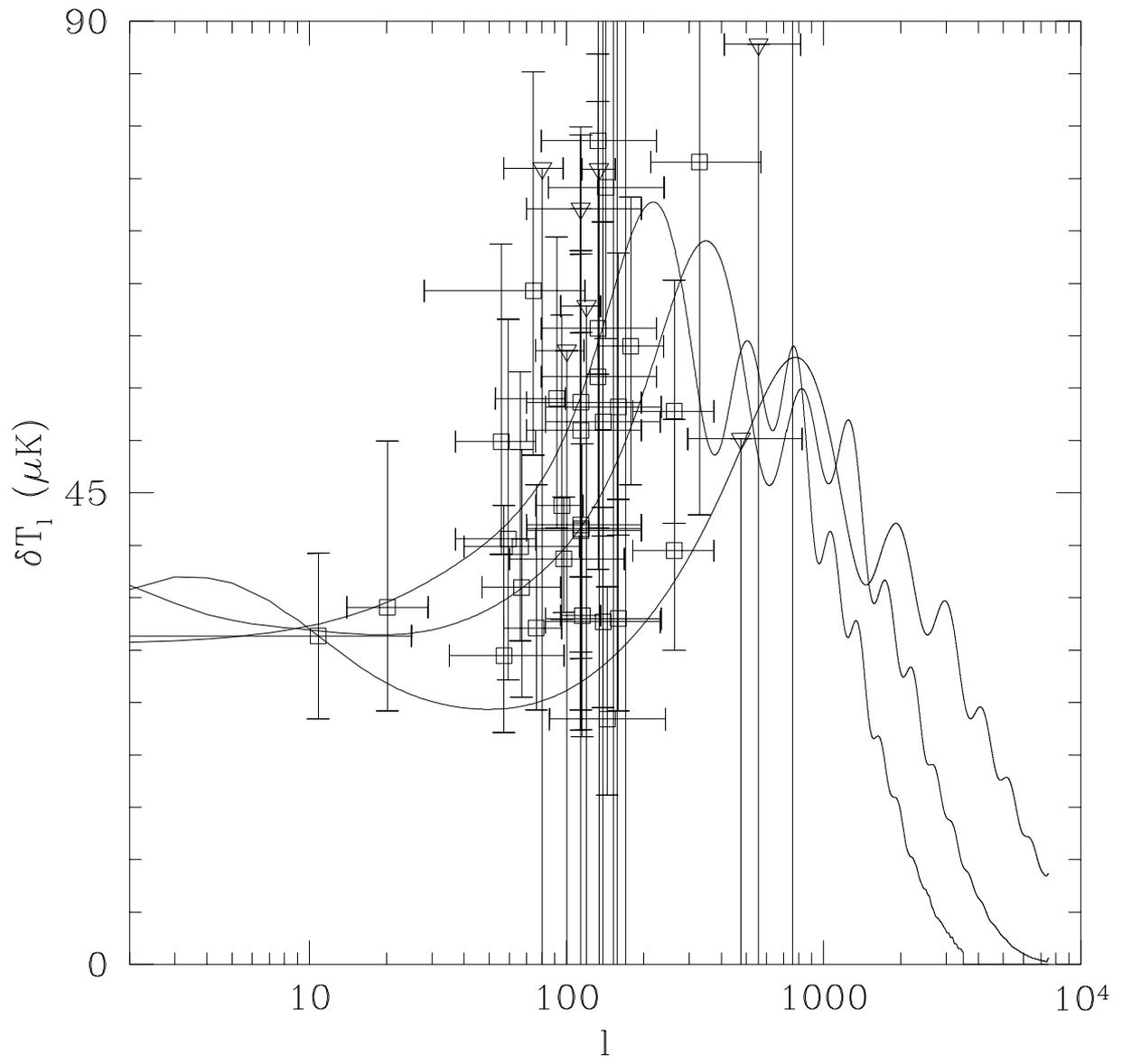

Figure 2